\documentclass[final,authoryear,5p,times,twocolumn]{elsarticle}
\usepackage{graphicx}
\usepackage{amssymb}
\usepackage{amsmath}

\usepackage{enumerate}
\usepackage{times}
\usepackage{jtb}
\usepackage{url}

\DeclareMathAlphabet{\bi}{OML}{cmm}{b}{it}

\journal{Journal of Theoretical Biology}

\begin{document}

\begin{frontmatter}

\title{The Joker effect: cooperation driven by destructive agents}

\author[tarraco]{Alex Arenas\corref{cor}}
\cortext[cor]{Corresponding author.} 
\author[bellaterra]{Juan Camacho}
\author[uc3m]{Jos\'e A. Cuesta}
\author[bellaterra]{Rub\'en Requejo}

\address[tarraco]{Departament d'Enginyeria Inform\`{a}tica i Matem\`{a}tiques,
Universitat Rovira i Virgili, 43007 Tarragona, Spain}
\address[bellaterra]{Department de F\'{i}sica,
Universitat Aut\'{o}noma de Barcelona, 08193 Bellaterra, Barcelona, Spain}
\address[uc3m]{Grupo Interdiciplinar de Sistemas Complejos (GISC),
Departamento de Matem\'{a}ticas,
Universidad Carlos III de Madrid, 28911 Legan\'{e}s, Madrid, Spain}

\begin{abstract}
Understanding the emergence of cooperation is a central issue
in evolutionary game theory. The hardest setup
for the attainment of cooperation in a population of individuals is
the Public Goods game in which cooperative agents generate a common
good at their own expenses, while defectors ``free-ride'' this good.
Eventually this causes the exhaustion of the good, a situation which is bad
for everybody. Previous results have shown that
introducing reputation, allowing for volunteer
participation, punishing defectors,
rewarding cooperators or structuring agents, can enhance cooperation.
Here we present a model which shows how the introduction of rare,
malicious agents ---that we term jokers--- performing just destructive
actions on the other agents induce bursts of cooperation. The appearance
of jokers promotes a rock-paper-scissors dynamics, where jokers outbeat
defectors and cooperators outperform jokers, which are subsequently
invaded by defectors. Thus, paradoxically, the existence of destructive
agents acting indiscriminately promotes cooperation.
\end{abstract}


\end{frontmatter}

\section{Introduction}
In the recent Hollywood movie \emph{The Dark Knight} (2008) the comic
character known as the Joker jeopardizes a whole society spreading
chaos and destruction with no aim of benefit at it. The situation is
so critical that even the mob is willing to cooperate with honest
people to stop this nonsensical catastrophe. This fiction provides
a visual metaphor of how an event like this can force exploiters of
society to collaborate temporarily to fight the common enemy. Society
is an emergent structure resulting from the cooperation among its 
members, and exploiters need society to survive, even if they do not
contribute to it. Thus they are specially sensitive to the destruction
of society precisely because, being selfish agents, society is their
only source of survival. The appearance of
the Joker provides a strong incentive for cooperation.

Beside situations like the one depicted by the Joker metaphor,
the importance of the inclusion of malicious agents on
the game is also illustrated in other scenarios.
Here are a few examples. Temporary coalitions
of rival parties are constantly formed whenever a common enemy arises,
only to restore their old rivalry once this enemy has been
wiped out. During the Second World War U.S.A.\ and U.S.S.R.\ were
allied in fighting Hitler, but they got engaged in the Cold War for
decades after the danger of Nazism had been ruled out. It is also well
known that strong affective links between humans are created when they
face a common difficult situation. Biology is another
source of potential examples. For instance, it has been shown that the 
perception of an increase in the risk of predation can induce cooperative
behavior in some bird species \citep{krams:2010}.
Indeed, prey species frequently form groups to 
increase the survival rate against predator attacks
\citep{hamilton:1971,krebs:1993}.
In some cases, this has been proven to happen even in the absence of
kinship among its members, as in the collective defense of spiny
lobsters \citep{lavalli:2009}.

The existence of these temporary coalitions for defense against a common
danger in rational and irrational agents alike calls for an evolutionary
explanation. In this article we propose a stylized evolutionary game
\citep{hofbauer:1998} aimed at studying theoretically this enhancement
of cooperation driven by the emergence of purely destructive agents. 
The game does not try to model any specific situation, but it proposes
an abstract setting in which the role of the indiscriminate destructive
action of these agents in enhancing cooperation is made clear. 
Our model is a modification
of the standard Public Goods (PG) game \citep{groves:1977}, the
$n$-players version of Prisoner's Dilemma and a paradigm of the risk
of exploitation faced by cooperative behavior \citep{hardin:1968}. 
It has been shown that several mechanisms involving reputation \citep{milinski:2005},
allowing for volunteer participation \citep{hauert:2002b,hauert:2002c}, punishing 
defectors \citep{fehr:1999,fehr:2000}, rewarding
cooperators \citep{sigmund:2001} or structuring agents
\citep{szabo:2002,wakano:2009,hauert:2008}, can enhance cooperation.
Here, we present a different mechanism 
for the enhancement of cooperation based on the existence of evil agents.
The game involves $n$
players who belong to one out of three different types: cooperators,
who contribute to the public good at a cost for themselves; defectors,
who free-ride the public good at no expense; and jokers, who do not
participate in the public good ---hence obtain no benefit whatsoever---
and only inflict damage to the public good. Groups are formed randomly, 
and each player's strategy is established before the group is selected.  
Hence, players have no memory. Remarkably, the appearance
of jokers promotes a rock-paper-scissors dynamics, where jokers outbeat
defectors and cooperators outperform jokers, which are subsequently
invaded by defectors. In contrast to previous models \citep{hauert:2002b,hauert:2002c},
the cycles induced by jokers are limit cycles, i.e. attractors of the dynamics,
and exist in the presence of mutations; these properties make them
robust evolutionary outcomes.
Therefore, paradoxically, the existence of destructive
agents acting indiscriminately promotes cooperation.


The paper is organized as follows. Section 2 exposes the model and shows the 
existence of cycles. Section 3 analyzes the dynamics for infinite populations, 
and section 4 compares the joker model with other RPS dynamics.

\section{A Public Good game with jokers: existence of limit cycles}

The PG game works as usual: every cooperator yields a benefit $b=rc$
($r>1$) to be shared by cooperators and defectors alike,
at a cost $c$ for herself (this cost can be set to $c=1$
without loss of generality: all other payoffs are given in units of
$c$), and defectors produce no benefit at all but get their share of the
public good. As for the new agents (jokers), every joker inflicts a
damage $-d < 0$ to be shared equally by all non-jokers
and gets no benefit. In a given game $0\le m \le n$ denotes the number of
cooperators, $0\le j\le n$ the number of jokers, and $n-m-j\ge 0$ 
the number of defectors; $S=n-j$ expresses the
number of non-jokers. In this group, the payoff of a defector will be
$\Pi_{\rm D}(m,j)= (rm-dj)/S$, and that of a cooperator
$\Pi_{\rm C} =\Pi_{\rm D}-1$. Then, in each group, defectors will always
do better than cooperators. Jokers' payoff is always $0$.

A usual requirement of PG games is that $r<n$. Without this requirement
the solution in which all $n$ players are defectors is no longer a Nash
equilibrium ---hence the dilemma goes away. As shown later, the evolutionary
dynamics for infinite populations yields the same constraint, i.e., if
$r<n$ the dynamics asymptotically approaches the tragedy of the commons.
However this is no longer true for finite populations, where the upper
bound of $r$ for which the tragedy of the commons takes place grows as $M$, 
the population size, decreases. In this case the
tragedy of the commons arises whenever $r<r_{\rm max}=n(M-1)/(M-n)$
(see~\ref{app:A}; notice in passing that for a population of $M=n$ individuals,
the evolutionary dynamics yields a tragedy of the commons for every $r>1$).

\begin{figure}
\begin{center}
\includegraphics[width=88mm]{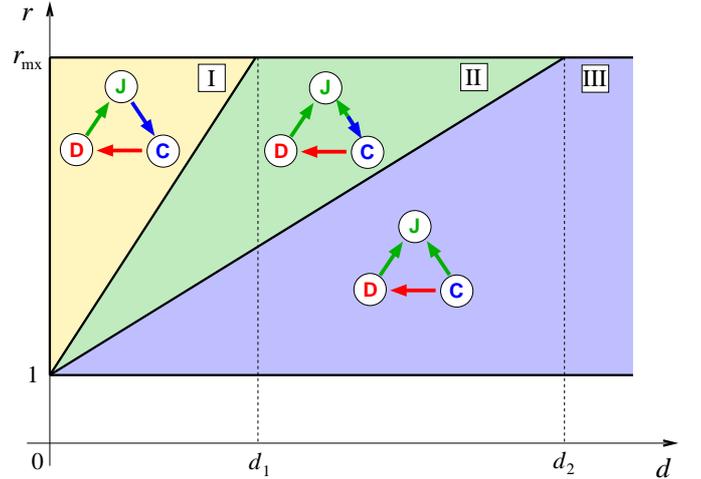}
\end{center}
\caption{{\bf Dynamics of invasions in a Public Goods game with jokers.} 
The axes represent the gain factor $r$ of the Public Goods game (i.e.,
the payoff each cooperator yields to the public good) and the ``damage''
$d>0$ that every joker inflicts on the public good. The tragedy of the commons
occurs for $1<r<r_{max}=n(M-1)/(M-n)$ (see text), which includes the dilemmatic region 
$1<r<n$ characteristic of PG games. Different colors are assigned to different 
invasion patterns: Light blue corresponds to a region
where J invades both C and D (III); light green corresponds to a region where
neither C nor J invades each other (there is bistability on the J--C line)
but D invades C and is in turn invaded by J, so again everything ends up
in J (II); finally, light yellow corresponds to a region where D invades C,
J invades D, but C invades J back, thus generating a rock-paper-scissors
cycle (I). The latter behavior is the essence of the \emph{Joker effect.}
The equations of the straight lines separating the three regions
are (from top to bottom) $r=1+d(n-1)$ and $r=1+d/(M-1)$. Notice that this
scheme is valid for arbitrary $n>1$. Also, for fixed $r$, all three
regions are crossed upon varying $d$, whereas vice versa is only true
provided $d<d_1=M/(M-n)$. The Joker effect does not occur if $d>d_1$.
For large populations, $M\gg 1$, the region for the rock-paper-scissors 
cycle simplifies to $n>r>1+(n-1)d$ and $d<1$.
\label{fig:diagram}}
\end{figure}

An invasion analysis provides the clue as to why a rock-paper-scissors
(RPS) cycle is to be expected when jokers intervene in the game. We
shall assume that we have a population of $M$ players of the same type
and will consider putative mutations of one individual to any of the
other two types. The mutation will thrive if the average payoff of the
mutant after many interactions overcomes the average payoff of a
non-mutant player. The result of this analysis (see~\ref{app:A}) is summarized
in Fig.~\ref{fig:diagram}, which represents the three different patterns
of invasion that can be observed within the region of interest
$1<r<r_{\rm max}$, $d>0$:
\begin{enumerate}[I.]
\item \textbf{Rock-paper-scissors cycle:} It arises whenever
$r>1+d(n-1)$. This condition expresses the fact that a single cooperator gets
a positive payoff in spite of the damage inflicted by $n-1$ jokers and
therefore being a cooperator pays (jokers get no payoff whatsoever).
\item \textbf{Joker-cooperator bistability:} If $1+d/(M-1)<r<1+(n-1)d$
neither jokers nor cooperators can invade each other. Nonetheless 
defectors always invade cooperators, and jokers always invade defectors,
so eventually only jokers survive, either because they are initially
a majority or indirectly through the emergence of defectors.
\item \textbf{Joker invasion:} If $r<1+d/(M-1)$ jokers will invade any
homogeneous population, so a homogeneous population of jokers is
the only stable solution. Notice that this region disappears for
large populations ($M\to\infty$) because $r>1$.
\end{enumerate}
The RPS cycle C$\to$D$\to$J$\to$C 
occurring in region I is the essence of the Joker effect.

\section{Infinite populations}
We can gain further insight into this effect by studying a replicator-mutator
dynamics \citep{maynard-smith:1982}. We assume a very large population in which
the three types are present at time $t$ in fractions $x$ (cooperators),
$y$ (defectors), and $z=1-x-z$ (jokers). Agents interact
with the whole population by engaging in the above described game within
groups of $n$ randomly chosen individuals \citep{hauert:2006}. Average
payoffs of a cooperator, a defector, and a joker are denoted $P_{\rm C}(x,z)$,
$P_{\rm D}(x,z)$, and $P_{\rm J}(x,z)$, respectively. Assuming individuals
of a given type mutate to any other type at a rate $\mu\ll 1$,
the replicator-mutator equations for this system will be
\begin{equation}
\begin{split}
\dot x &= x (P_{\rm C} - \bar{P}) + \mu (1-3x) ,\\
\dot y &= y (P_{\rm D} - \bar{P}) + \mu (1-3y), \\
\dot z &= z (P_{\rm J} - \bar{P}) + \mu (1-3z),
\end{split}
\label{eq:RM}
\end{equation}
where $\bar{P}=xP_{\rm C}+ yP_{\rm D}+zP_{\rm J}$ is the mean payoff
of the population at a given time. Explicit expressions for $P_{\rm C}$,
$P_{\rm D}$, and $P_{\rm J}$ can be obtained by averaging over all
samples of groups of $n$ players extracted from a population containing
$Mx$ cooperators, $My$ defectors, and $Mz$ jokers, in the limit of very
large populations ($M\to\infty$); the derivation can be found in~\ref{app:B}.
Let us recall that the parameters of the game in the infinite population limit satisfy $1 < r < n$ and $d > 0$; 
the first condition enforces the public goods dilemma,
and the second one implies that jokers beat defectors in the absence of cooperators, because
defectors receive the damage inflicted by jokers thus obtaining a negative payoff.

\begin{figure}
\begin{center}
\includegraphics[width=88mm]{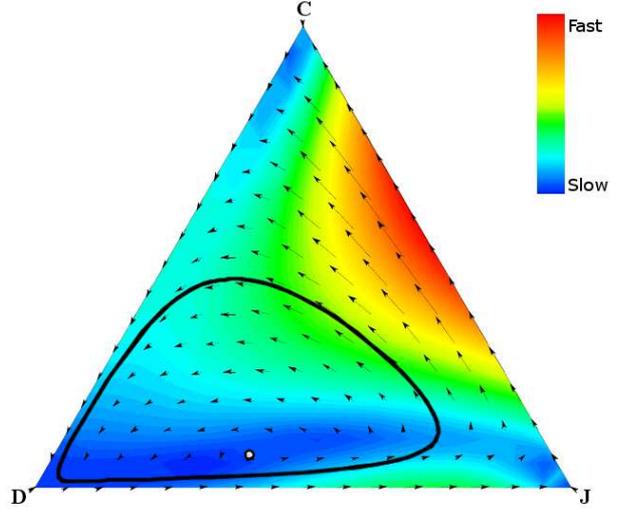}
\end{center}
\caption{{\bf The Joker effect in public goods games for large, well-mixed
populations.} The simplex describes the replicator-mutator dynamics,
Eq.~\ref{eq:RM}, for a population of cooperators, defectors and jokers
with parameter values satisfying $n>r>1+d(n-1)$, for which a rock-paper-scissor
dynamics is expected (yellow region in Fig.~\ref{fig:diagram}). When mutation
rates are small, the only equilibrium is a repeller (white dot), and
trajectories end up in a stable limit cycle (black line). Thus the presence
of jokers induces periodically a burst of cooperators. Cooperators abound
during short time spans, as shown by the small fraction of cooperators in
the equilibrium point. 
Parameters: $n=5$, $r=3$, $d=0.4$ and $\mu=0.005$. 
(Images generated using a modified version of the Dynamo Package
\citep{sandholm:2007}).
\label{fig:simplex}}
\end{figure}

The stability analysis of the dynamical system (\ref{eq:RM}) recovers the
picture displayed in Fig.~\ref{fig:diagram} (taking $M\to\infty)$. When
$r<1+(n-1)d$ the system is in region II. The only stable equilibrium is
a population of only jokers and any trajectory of (\ref{eq:RM})
is asymptotically attracted to it. Thus, in this region the destructive
power of jokers is high enough to wipe out the populations of both 
cooperators and defectors. But the most interesting situation takes
place when 
\begin{equation}
\label{ineq}
r>1+(n-1)d, 
\end{equation}
i.e., in region I. In the absence of mutations the dynamical system (\ref{eq:RM}) 
has three saddle points at the corners of the simplex
as well as an unstable mixed equilibrium (see~\ref{app:C}). As a
consequence, the attractor of the system is the heteroclinic orbit 
${\rm C}\to {\rm D}\to {\rm J}\to {\rm C}$. The period is infinite because 
the system delays more and more around the saddle points.
When mutations occur the corners of the simplex are no longer equilibria,
and one is left with the interior fixed point, which for small mutations is
a repeller (see~\ref{app:C}). Since trajectories are confined within the closed
region of the simplex, they are attracted to a stable limit cycle for
any $r>1$ (a direct consequence of the Poincar\'e-Bendixon
theorem \citep{simmons:2006}), as shown in Fig.~\ref{fig:simplex}.

\begin{figure}
\centering
\includegraphics[width=88mm]{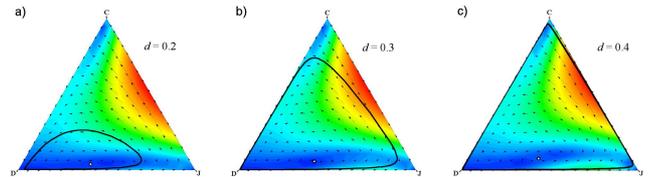}
\caption{{\bf Replicator-mutator dynamics as a function of the damage $d$ inflicted by jokers}.
For a fixed mutation rate, the size of the cycles increases as the damage increases.
Parameters: $n=5$, $r=3$ and $\mu=0.001$. 
\label{FigureSI1}}
\end{figure}

\begin{figure}
\centering
\includegraphics[width=88mm]{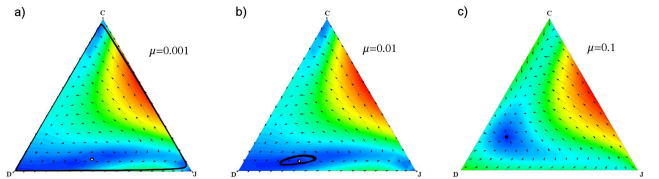}
\caption{{\bf Replicator-mutator dynamics as a function of the mutation rate
$\mu$}. (a) For very small mutation rates cycles approach the boundary of the
simplex. (b) As $\mu$ increases, the cycle amplitude decreases and, above a
critical value (typically, $\mu_c \simeq 0.01$), cycles disappear in a Hopf
bifurcation yielding a stable mixed equilibrium (c). 
Parameters: $n=5$, $r=3$ and $d=0.4$. 
\label{FigureSI2}}
\end{figure}

The size of the cycle depends on the parameter values. It grows as $d$ increases
---i.e., when jokers play a more important role
(Fig.~\ref{FigureSI1})--- and as the mutation rate
decreases (Fig.~\ref{FigureSI2}). For both, large values of $d$ [compatible with condition \eqref{ineq}]
and very small mutations, the cycle closely follows the boundaries of the
simplex (see Fig.~\ref{FigureSI2}a). By increasing the mutation rate (typically over $0.01$), 
cycles disappear in a Hopf bifurcation yielding a stable mixed equilibrium (Figs.~\ref{FigureSI2}b-c).

\section{Discussion and conclusions}

This evolution has some resemblances with the effect of volunteering in
a PG game \citep{hauert:2002b,hauert:2002c}, but the two games are fundamentally
different. This can be told from the dynamic behavior of the system.
In both cases, the existence of a third agent which does not participate in
the game is the ultimate reason why cooperators periodically thrive
through a Rock-Paper-Scissor dynamics. However, while the loners game
leads to neutrally stable cycles around a center, trajectories in the
Joker model are attracted by the heteroclinic cycle C--D--J--C. The
difference is even more striking if mutations are included. Mutations
replace the cycles in the loner model by a stable mixed equilibrium.
In contrast, in the Joker model mutations substitute the heteroclinic
orbit by a stable limit cycle, which undergoes a transition (Hopf
bifurcation) to a stable mixed equilibrium above a threshold mutation rate.

These two scenarios can be understood from the analysis of general RPS
games \citep{hofbauer:1998}. There are three situations: (a) orbits are
attracted towards an asymptotically stable mixed equilibrium (the case
of the loners game with mutations), (b) orbits cycle around a neutrally
stable mixed equilibrium (the case of the loners game without mutations),
and (c) orbits go away from an unstable mixed equilibrium and approach
the heteroclinic orbit defined by the border of the simplex (the case
of the Joker game without mutations). If mutations are added to the
latter type of RPS games, limit cycles and a Hopf bifurcation upon
increasing the mutation rate are also found \citep{mobilia:2010}.
Limit cycles are robust to perturbations and have a well defined amplitude
irrespective of the initial fractions of players (as long as it is
not at the border of the simplex). Therefore, they are true attractors of
the dynamics, and can thus be regarded as a robust evolutionary outcome, 
in contrast to neutrally stable cycles. 

\begin{figure}
\centering
\includegraphics[width=88mm]{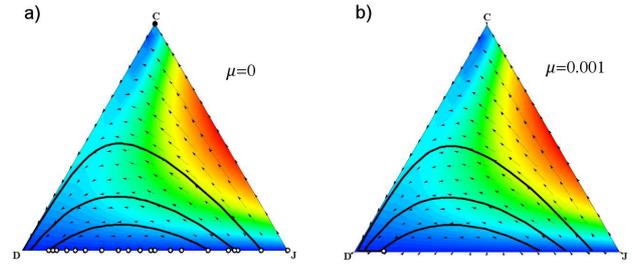}
\caption{{\bf Replicator-mutator dynamics for $d=0$}. If jokers are just
passive agents cooperators go extinct. (a) $\mu=0$. The system ends up in a
point of the line DJ with a majority of defectors. (b) $\mu=0.001$. Mutation
generates one single stable state made up mostly by defectors. 
Parameters: $n=5$, $r=3$ and $d=0$. 
\label{FigureSI3}}
\end{figure}

In contrast to loners, which do not participate in the game but receive
a benefit outside of it, jokers do not receive any benefit at all and
cause damage to players. Both loner and joker models coincide ---in
the absence of mutations--- when the damage inflicted by jokers and
the benefit obtained by loners are both zero. In this case both become
simply non-participants in the game, and the only effect they produce
is a reduction in the effective number of players in the game, which
is not enough to induce an oscillatory dynamics (see Fig.~\ref{FigureSI3}).
In other words, the appearance of the RPS cycle which periodically
increases the population of cooperators in the presence of jokers can
only happen, remarkably, provided $d>0$, i.e., if jokers are truly
destructive agents. 

In this letter we have shed light on a still unexplored aspect of evolutionary
game theory (the presence of a destructive strategy) in the prototypical
PG game. We have shown, both theoretically and by numerical
simulations, that the addition of purely destructive agents (jokers)
to a standard PG game has, paradoxically, a positive effect on cooperation.
Bursts of cooperators are induced through the appearance of a RPS cycle in
which jokers beat defectors, who beat cooperators, who beat jokers in
succession.
The evolutionary dynamics provoked by the Joker, with periods of cooperation,
defection and destruction of the PG, may help understand the appearance of
cognitive abilities that allow individuals to foresee the destructive periods,
promoting in advance the necessary cooperation to avoid them.

We have proven this ``Joker effect'' to occur both in
finite and infinite populations, discarding the possibility of its being
an artificial size-depending phenomenon. Further research is required to ascertain 
the scope of the constructive role of destruction in general settings. 
This provides a new framework for
the evolution of cooperation that may find important implications in social,
biological, economical, and even philosophical contexts, and that is worth
exploring either with different variants of this game or with new, more
specific games accounting for indiscriminate destruction.

\section*{Acknowledgments}
Financial support from Ministerio de Ciencia y Tecnolog\'{\i}a
(Spain) under projects FIS2009-13730-C02-02 (A.A.), FIS2009-13370-C02-01
(J.C. and R.J.R.) and MOSAICO (J.A.C.); from the Director, Office of Science,
Computational and Technology Research, U.S. Department of Energy under
Contract No.\ DE-AC02-05CH11231 (A.A.); from the Barcelona Graduate School of Economics and of the Government of Catalonia (A.A.); from the Generalitat de Catalunya under project 2009SGR0838 (A.A.)
2009SGR0164 (J.C. and R.J.R.) and from Comunidad de Madrid under project MODELICO-CM
(J.A.C.). R.J.R. acknowledges the financial support of the Universitat
Aut\`onoma de Barcelona (PIF grant) and the Spanish government (FPU grant).

\appendix

\section{Finite populations: invasion analysis}
\label{app:A}

We shall consider the situation in which in a homogeneous population
of $M$ individuals with the same strategy Y, one of them mutates
(changes) to a different type X.
The new individual will invade
provided its average payoff after many interactions, $P_{\rm X}$,
is larger than the average payoff of a Y individual, i.e.,
$P_{\rm X}>P_{\rm Y}$. Average payoffs can be evaluated as follows.
The population is made of one X player and $M-1$ Y players. Thus,
when playing the game, the X player will always interact with
$n-1$ Y players. Therefore
\begin{equation}
P_{\rm X}=\Pi_{\rm X}(1X,(n-1)Y).
\end{equation}
On the other hand, the $n-1$ opponents of a Y player can be of just
two types: either all $n-1$ are Y players, or $n-2$ are Y players
and one is the single X player. The latter situation occurs with
probability $(n-1)/(M-1)$. Therefore the average payoff of a Y player
will be
\begin{equation}
P_{\rm Y}=\Pi_{\rm Y}(nY)\frac{M-n}{M-1}+\Pi_{\rm Y}(1X,(n-1)Y)\frac{n-1}{M-1}.
\end{equation}
Next we derive the invasion conditions for homogeneous populations
of three types of players. In this new scenario we must consider the
six different situations 
arising form the pair interactions that can be formed:
\begin{enumerate}[(A)]
\item {${1{\rm D}+(M-1){\rm C}}$.}
\begin{equation}
P_{\rm C}=r-1-\frac{r}{n}\frac{n-1}{M-1}, \qquad
P_{\rm D}=r-\frac{r}{n}.
\end{equation}
The tragedy of the commons occurs when defectors overcome cooperators,
i.e., $P_{\rm D}>P_{\rm C}$. This happens iff
\begin{equation}
r<n\frac{M-1}{M-n}.
\label{eq:rmax} 
\end{equation}
We will henceforth assume \eqref{eq:rmax} to hold. This condition contains
the dilemmatic region $1<r<n$ of PG games. In the limit $M\to\infty$, the
inequality \eqref{eq:rmax} reduces to $r<n$ and both, the conditions for
the dilemma and the tragedy of the commons coincide.

\item{${1{\rm C}+(M-1){\rm D}}$.}
\begin{equation}
P_{\rm C}=\frac{r}{n}-1, \qquad
P_{\rm D}=\frac{r}{n}\frac{n-1}{M-1}.
\end{equation}
Because of \eqref{eq:rmax} $P_{\rm D}>P_{\rm C}$, so \emph{C never invades D.}

\item{${1{\rm J}+(M-1){\rm C}}$.}
\begin{equation}
P_{\rm C}=r-1-\frac{d}{M-1}, \qquad
P_{\rm J}=0.
\end{equation}
Since $P_{\rm J}>P_{\rm C}$ iff 
\begin{equation}
r<1+\frac{d}{M-1},
\label{eq:J-C}
\end{equation}
then \emph{J invades C iff \eqref{eq:J-C} holds}.

\item{${1{\rm C}+(M-1){\rm J}}$.}
\begin{equation}
P_{\rm C}=r-(n-1)d-1, \qquad
P_{\rm J}=0.
\end{equation}
Since $P_{\rm C}>P_{\rm J}$ iff 
\begin{equation}
r>1+(n-1)d,
\label{eq:C-J}
\end{equation}
then \emph{C invades J iff \eqref{eq:C-J} holds}.

\item{${1{\rm D}+(M-1){\rm J}}$.}
\begin{equation}
P_{\rm D}=-(n-1)d, \qquad
P_{\rm J}=0.
\end{equation}
As long as $d>0$ we will have $P_{\rm J}>P_{\rm D}$, then
\emph{D never invades J}.

\item{${1{\rm J}+(M-1){\rm D}}$.}
\begin{equation}
P_{\rm D}=-\frac{d}{M-1}, \qquad
P_{\rm J}=0.
\end{equation}
As long as $d>0$ we will have $P_{\rm J}>P_{\rm D}$, then
\emph{J always invades D}.
\end{enumerate}

Figure~\ref{fig:diagram} illustrates the different regions of interest 
in this game. The most interesting one is that in which there is a 
rock-paper-scissor rotation between C, D, and J, which corresponds to
\begin{equation}
1<r<n\frac{M-1}{M-n}, \qquad 0<d<\frac{r-1}{n-1}.
\end{equation}

\section{Infinite populations: average payoffs}
\label{app:B}

We evaluate here the average payoffs $P_{\rm X}$ obtained by each strategy
($i=$C, D, J) in this game when the population is very large. These functions
will determine the dynamics of the population through the replicator equation.
As before, sample groups of $n$ individuals playing the game are randomly formed,
and it is assumed that each player is sampled a large number of times before
payoffs are compared in order to update strategies. The payoff for a given
strategy is therefore proportional to the average payoff that a player using
this strategy obtains playing against the whole population. This average
payoff will depend only on the player's strategy and the composition of the
population, described by a fraction $x$ of cooperators, $z$ of jokers and
$y=1-x-z$ of defectors. Notice that $P_{\rm J}=0$ for any composition of the
population, so only cooperators' and defectors' payoffs need to be calculated.

\subsection{Defectors}

The average payoff of a defector is  
\begin{equation}
P_D=\left\langle\frac{rm-dj}{S}\right\rangle,
\label{PD1}
\end{equation}
where the symbol $\langle\cdots\rangle$ denotes an average
over samples of
$n-1$ opponents randomly selected from the population. The average $\langle
m/S\rangle$ can be obtained as in \citep{hauert:2002c}, yielding
\[
\left\langle\frac{m}{S}\right\rangle=\frac{x}{1-z}\left(1-\frac{1-z^n}{n(1-z)}\right).
\]
Since $j=n-S$, the second term in Eq.~\eqref{PD1} can be written as
$n\langle 1/S\rangle-1$, where
\[
\left\langle\frac{1}{S}\right\rangle=\sum_{S=1}^n \binom{n-1}{S-1}
(1-z)^{S-1}z^{n-S}\frac{1}{S},
\]
the factor in front of $1/S$ in the summation being the probability of having
$S-1$ non-jokers in a group of $n-1$ randomly chosen players. By using the
identity $a\binom{a-1}{b-1}=b\binom{a}{b}$, the latter expression becomes
\[
\left\langle\frac{1}{S}\right\rangle=\frac{1-z^n}{n(1-z)}.
\]
Joining the two averages one gets the average payoff of a defector,
\begin{equation}
P_{\rm D}= r\frac{x}{1-z}\left(1-\frac{1-z^n}{n(1-z)}\right)-d
\left(\frac{1-z^n}{1-z}-1\right),
\label{PD}
\end{equation}
the first term arising from the exploitation of cooperators and the second
one being the damage inflicted by jokers.

\subsection{Cooperators}

The difference $P_{\rm D}-P_{\rm C}$ can be written as 
\begin{equation}
P_{\rm D}-P_{\rm C}=\left\langle 1-\frac{r}{S}\right\rangle
\label{dif1}
\end{equation}
because in a group of $S-1$ opponents switching from cooperation to
defection yields a payoff increment of $1-r/S$: the defector's payoff gets
reduced by $r/S$ because there is one cooperator less in the group, but adds
$1$ to her payoff because she does not pay the cost of cooperating
\citep{hauert:2002c}. The average in the r.h.s.\ of Eq.~\eqref{dif1} just
contains $<1/S>$, thus yielding
\begin{equation}
P_{\rm D}-P_{\rm C}=1-\frac{r}{n}\frac{1-z^n}{1-z}.
\label{dif}
\end{equation}
Finally, from Eqs.~\eqref{PD} and \eqref{dif} one gets
\begin{equation}
\begin{split}
P_C =&\,r\frac{x}{1-z}\left(1-\frac{1-z^n}{n(1-z)}\right)+
\frac{r}{n}\frac{1-z^n}{1-z}-1 \\
& -d \left(\frac{1-z^n}{1-z}-1\right).
\end{split}
\label{PC}
\end{equation} 

\section{Infinite populations: proof of existence of limit cycles}
\label{app:C}

To complete the proof that the system ends up in a limit cycle it remains to
show that the interior equilibrium of Eqs~(\ref{eq:RM}) is a repeller, i.e., its
two eigenvalues have positive real parts. The interior equilibrium and its
stability can be evaluated in the limit of small mutation rates, the one we
are interested in. In this case, one can neglect the dependence of $\mu$ in the
position of the fixed point. We are thus faced with the solution of the
dynamical system (\ref{eq:RM})  without the mutation term. The calculation becomes
simple for $n=2$, and tractable for $n>3$. The proofs are treated separately
in the next subsections.

\subsection{Interior fixed point for $n=2$}

The interior fixed point $(x_0,y_0,z_0)$ satisfies $P_{\rm C}=P_{\rm D}=0$.
According to Eq.~\ref{dif}, the first equality requires $(1+z_0)r=2$,
yielding
\[
z_0=\frac{2-r}{r}.
\] 
Since $n=2>r>1$, one has $0<z_0<1$, as it should. The second equality, $P_{\rm D}=0$, produces
\[
x_0=2d\,\frac{2-r}{r^2}.
\]
Condition $r>1+d$ from expression \eqref{ineq} guarantees that $0<x_0<1$ and
$0<y_0=1-x_0-z_0<1$.
In order to analyze the stability of this equilibrium, we consider frequencies
$x$ and $z$ as the independent variables of the two-dimensional system. To
prove that the equilibrium is a repeller it suffices to show that the trace
and determinant of the Jacobian matrix at the fixed point are both positive.
For $n=2$, equations~(\ref{eq:RM}) become
\begin{align}
\dot x &= -\frac{1}{2}x(2dz^2-r z+2-r-2x+r x), \\
\dot z &= z[(1-r)x + dz(1-z)].
\end{align}
The Jacobian matrix in the interior equilibrium is
\begin{equation}
\begin{pmatrix}
\displaystyle \frac{d(2-r)^2}{r^2}  &
\displaystyle \frac{d(2-r)(r^2+4dr-8d)}{r^3} \\[4mm]
\displaystyle -\frac{(2-r)(r-1)}{r} &
\displaystyle \frac{d(2-r)(3r-4)}{r^2}
\end{pmatrix},
\end{equation}
whose trace, $T$, and determinant, $D$, are
\begin{align}
T &=\frac{2d(2-r)(r-1)}{r^2}>0, \\
D &=\frac{d(r-2)^2(r^2+r(d-1)-2d)}{r^3}>0.
\label{tracedet}
\end{align}
$T$ is positive because $n=2>r>1$. To prove that the determinant is positive,
we should realize that the second bracket in its expression can be written as
$r(r-1)-d(2-r)$, which is larger than $2(r-1)^2>0$ because $r>1+d$. 

\subsection{Interior fixed point for $n>3$}

We use the same procedure as in the previous case. The fraction of jokers
$z_0$ of the interior equilibrium arises from $P_{\rm C}=P_{\rm D}$, namely
Eq.~\eqref{dif}. Once it is found, $x_0$ follows from $P_{\rm D}=0$,
c.f.\ Eq.~\eqref{PD}.

\subsubsection{Calculation of $z_0$}

$z_0$ is obtained as the solution to
\begin{equation}
1-\frac{r}{n}\frac{1-z^n}{1-z}=0,
\label{z0}
\end{equation}
which is equivalent to  
\begin{equation}
\sum_{i=0}^{n-1} z^i=n/r.
\label{z0bis}
\end{equation}
The latter equation has exactly one solution, namely the crossing of the
polynomial in the l.h.s of Eq.~\eqref{z0bis} with the constant $n/r>1$. 
Since $r>1$, this occurs at $0<z_0<1$, consistent with
the meaning of $z_0$. There is no analytical solution to Eq.~\eqref{z0} for
arbitrary $n$. There exists, however, a simple analytical solution in the
limit of large $n$, which is indeed an excellent approximation for all
$n>3$. It can be obtained neglecting $z^n$ as compared to 1 in \eqref{z0}, which leads to
\begin{equation}
z_0\approx 1-\frac{r}{n}.
\label{z0tris}	
\end{equation}
Since $r<n$, one has, of course, $0<z_0<1$. For consistence,
$z_0^n=(1-\frac{r}{n})^n\approx e^{-r}\ll 1$, 
which holds, say, for $r> 3$. Notice that if $r\ll n$ the
equilibrium approaches allJ, so that cycles get very close to this
state in this limit.

\subsubsection{Calculation of $x_0$}

Let us impose $P_{\rm D}=0$. Introducing \eqref{z0} into \eqref{PD} one finds
\begin{equation}
x_0\approx \frac{d}{r-1}\left(\frac{n}{r}-1\right)(1-z_0).
\label{x0}
\end{equation}
Conditions $d>0$, $n>r>1$, and $r>1+(n-1)d$ yield $0<x_0<1$ and $0<x_0+z_0<1$,
so that the three fractions are smaller than $1$. Substituting
$z_0$ from expression \eqref{z0tris} into \eqref{x0} one finally obtains
\begin{equation}
x_0 \approx \frac{d}{r-1}\left(1-\frac{r}{n}\right).
\label{x0bis}	
\end{equation}

\subsubsection{Stability of the interior equilibrium}

We need to determine the Jacobian matrix for the equilibrium $(x_0,z_0)$
given by \ref{x0} and \ref{z0tris}. The dynamical system (\ref{eq:RM})  can be
written as
\begin{align}
\dot x =&-\frac{x}{n(1-z)^2}\Big(-r+n-rnxz-dnz^{n+1}+2nxz \nonumber \\
&-nxz^2 -2nz+rx-rxz^n +dnz^2-nx+rz^n \\
&+nz^2+rz-rz^{n+1}+rnxz^2+dnz^{n+2}-dnz^3\Big), \nonumber \\
\dot z =& -(-dz+rx+dz^n-x)z.
\end{align}
The first equation is very cumbersome. Fortunately, as already explained,
in the limit of large $n$ and if $r> 3$ one can neglect terms of
order $z^n$ and above. Using expressions \eqref{z0tris} and \eqref{x0bis},
the Jacobian matrix $J$ can be written as $J=Yd(n-r)/n$, where
\begin{equation}
Y=
\begin{pmatrix}
\displaystyle \frac{n-r}{r}  &
\displaystyle \frac{nr(r-1)+d(r-n)(r^2-r+n)}{r^2(r-1)^2} \\[4mm]
\displaystyle -\frac{r-1}{d} &
\displaystyle 2
\end{pmatrix}.
\end{equation}
(Notice that the factor $d(n-r)/n>0$.)
As the diagonal elements of this matrix are positive, the trace is positive.
Also $Y_{zx}<0$ and, as we show next, $Y_{xz}>0$, therefore the determinant
turns out to be positive, and the interior equilibrium is a repeller. To see that
$Y_{xz}>0$ we must show that the numerator is positive. This can be
shown by writing it as
\[
nr(r-1)+d(r-n)(r^2-r+n)> (r-1)^2\frac{(n-r)^2+nr}{n-1}>0.
\]
The first inequality follows from condition $r-1>(n-1)d$.

\bibliographystyle{model5-names}
\bibliography{evol-coop}


\end{document}